\documentstyle[aps]{revtex}

\newcommand\BEQ{\begin{equation}}
\newcommand\EEQ{\end{equation}}
\newcommand\BEQB{\begin{eqnarray}}
\newcommand\BEQA{\begin{eqnarray*}}
\newcommand\EEQB{\end{eqnarray}}
\newcommand\EEQA{\end{eqnarray*}}
\newcommand\BF{\begin{figure}}
\newcommand\EF{\end{figure}}

\newcommand{\degc}{${}^{\circ}$C}

\begin{document}
\input{epsf}

\title{Discerning Aggregation in Homogeneous Ensembles:\\
A General Description of Photon Counting Spectroscopy in Diffusing Systems}

\author{Hai-Cang Ren, Noel L. Goddard,$^{\ast}$  Gr\'egoire 
Bonnet,$^{\ast} {\dagger}$ and Albert 
Libchaber$^{\ast}$ }
\address{Laboratory for Theoretical Physics, The Rockefeller 
University, New York, NY 10021 \\ $^{\ast}$Center for Studies in 
Physics and Biology, The Rockefeller University, 
New York, NY 10021\\ ${\dagger}$ Present Address: Laboratory of Immunology, 
National Institute of Allergies and Infectious Diseases\\ 
National Institute of Health, Bethesda MD 20892}
\
\date{\today}
\setlength{\topmargin}{0.1cm}
\maketitle

\renewcommand{\baselinestretch}{2}\normalsize
\begin{abstract}
In order to discern aggregation in solution, we present a quantum 
mechanical analogue of the photon statistics from fluorescent 
molecules diffusing through a focused beam.  A novel generating functional is 
developed to fully describe the experimental physical system, as well 
as the statistics.  Histograms of the measured time delay between photon 
 counts are fit by an analytical solution describing the static as 
well as diffusing regimes.  To determine empirical fitting parameters, 
fluorescence correlation spectroscopy is used in parallel to the photon 
counting.  For expediant analysis, we find that the distribution's deviation 
from a single Poisson shows a difference between two 
single fluor monomers or a double fluor aggregate of the same total intensities.  
Initial studies were preformed on fixed-state aggregates limited to 
dimerization.  However preliminary results on reactive species 
suggest that the method can be used to characterize any aggregating system.
\end{abstract}

\vspace{0.3cm}


\vspace{0.5cm}

\section{Introduction}
Aggregation and cooperative binding are fundamental to biological 
function 
and regulation, but difficult to observe at the few molecule level.  
Recent advances in few molecule solution spectroscopy have been achieved 
by combining the comparitively large signals of fluorescence with 
recent technological advances in photon counting (e.g. low noise 
detectors).  
One powerful technique, fluorescence correlation spectroscopy (FCS), 
has enabled researchers to observe many small 
ensemble processes such as diffusion \cite{Magde72}, 
molecular conformational dynamics \cite{Bonnet98}, and reaction 
kinetics \cite{Elson74}.  In FCS, fluctuations in 
fluorescence intensity are temporally correlated to reveal the 
timescale of the underlying fluctuation 
source.  In the simplest case, this could be the 
diffusion timescale of a fluorescent molecule through a sampling 
volume.  However, it is difficult to discern aggregation based on the diffusion 
time alone since the increase in diffusion time between a monomer or a dimer is 
weakly dependent on the increase in effective radius 
($\propto  <r^{2}>^{1/2}$ for hard spheres) \cite{CantorandSchimmel}.   

\begin{figure}
\centerline{\epsfxsize 6in \epsfbox{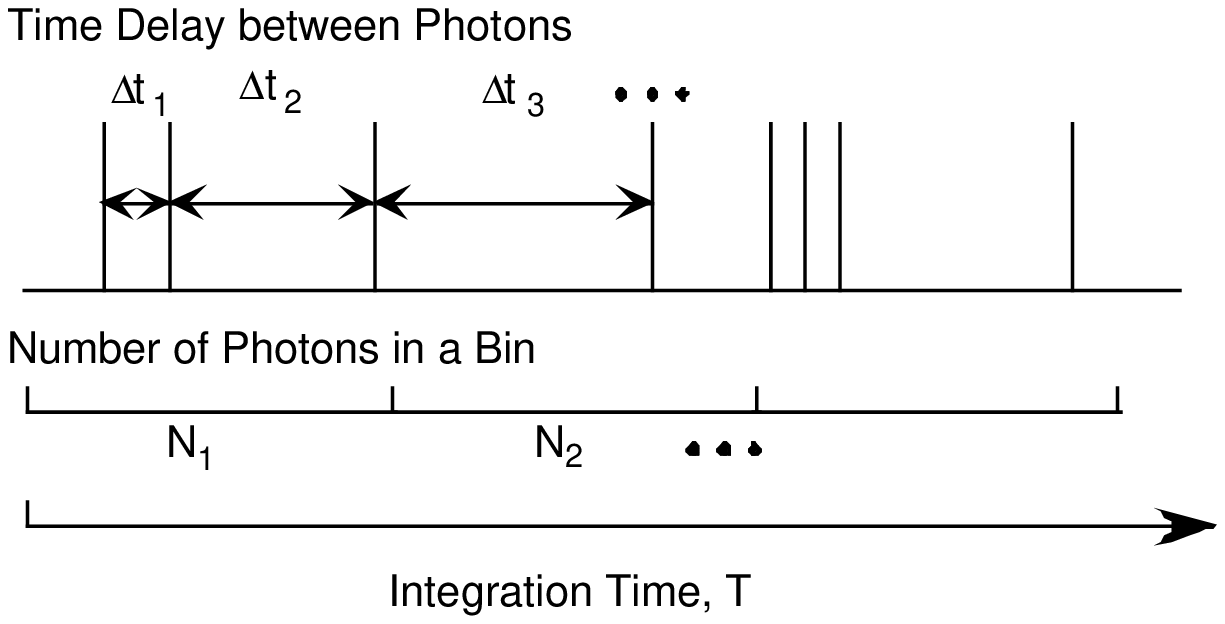}}
\vskip0.3cm
\caption{}
\label{fig:counting}
\end{figure}

An alternative approach to FCS is the statistical analysis of the time series of 
emitted photons (number counting Fig \ref{fig:counting}), since the number 
distribution of photons in the 
detection volume at any given moment is different for species of 
different quantum yield or fluorophore number.  This analysis, called 
photon number counting histogram analysis \cite{Chen99,Kask99} was originally developed 
to detect the ``brightness'' per particle, but relied on the static (non-diffusing) 
limit.  However in experiment, several factors potentially overshadow the specific brightness signature 
of a fluor, such as triplet states, bleaching, or quenching. 
We propose a new method of analysis, the {\em time delay histogram}, to discern small differences 
in a specie's fluorescence complemented by differences in diffusive behavior. The time delay 
histogram is constructed of the time between successive photon 
counts (Fig \ref{fig:counting}).  Unlike counting the number of events per 
time bin, this type of counting allows us to extract additional information 
about the diffusion.  Conceptually, as more fluorinated monomers aggregate, the 
fluorophores sample the detection volume in bunches. 
This increases the chance of a large 
time interval between two successive emission photons and the large $\Delta t$ tail of 
the time delay histogram has an increased frequency. 

To extract the information on the structural change, we have developed a generating functional to unify 
different statistical aspects of the photon 
time series. This functional is modeled as a transition matrix element 
of a fictitious quantum mechanical system with time variable continuated to the imaginary axis. 
Many well developed techniques in quantum 
mechanics are borrowed to derive analytical expressions for other  
experimental observables such as the auto correlation function.  Our approach 
is complementary to the description model 
proposed by Novikov and Boens \cite{Novikov01} for the photon 
counting histogram.  An extension of our functional can also be applied to modern multi-channel 
techniques \cite{Schwille97,Kask00} as well as the incorporation of secondary 
processes such as triplet states, bleaching, quenching, and chemical reaction 
kinetics.

Recently Kask, {\em et.al} also proposed a method to extract the diffusion signature of 
a particle through fluorescence intensity multiple distribution analysis 
(FIMDA) \cite{Kask00}. In FIMDA, many 
histograms are constructed from the same time trace as the bin size is varied.  
In contrast, the time delay histogram we propose, contains 
the diffusion signature of the aggregate in a 
{\em single} histogram.  At small time delays (less than the diffusion 
time through the collection volume) we are sensitive to the rate of photons emitted per object, and at 
greater delays, diffusion dominates the statistics. 
Finally, we offer a complementary technique to FIMDA using the Mandel 
Q parameter [1] to expediantly 
extract diffusion information from number counting distributions.  
The power of this technique could 
be increased by greater photon collection 
capabilities, leading to better statistics and ultimately better 
discrimination, including higher moments, in the large delay limit.

To demonstrate the sensitivity of our analysis, we choose a  
case where the difference in diffusion times are negligible.  A particularly 
difficult case is the discrimination of two monomers of identical
fluorescence  from a single dimer with the sum of their fluorescence.  Although 
both have the same average rate of emitted photons, they sample 
the excitation beam profile differently as they diffuse 
(inhomogeneous intensity profile of a focused beam).  For the 
experiment, a sequence of single stranded DNA is specifically tagged 
with either one or two fluorophores per strand.  The single dyed 
strands will be considered ``monomers'' and the doubled tagged as 
``dimers''.   For all cases, we find good 
discrimination between samples of a given concentration of dimers 
versus that of twice the concentration of monomers, where 
both samples have the same average fluorescence.

\section{Overview of Approach}
 
The time series of photon 
events may be described by the instantaneous fluorescence intensity,
$$I(t)=\sum_n^N\delta_\epsilon(t-t_n)\eqno(2.1)$$ where 
$t_1, t_2,...,t_n,...t_N$ corresponds 
to the tick marks of Fig.1 with $N$ total number of photons counted and $\epsilon$ is the detector resolution.  
$\delta_\epsilon(t-t_n)={1\over\epsilon}$ for 
$|t-t_n|<\epsilon$ and $\delta_\epsilon(t-t_n)=0$ otherwise. 
In the theoretical analysis of the subsequent sections, we shall 
take the limit $\epsilon\to 0$ so the 
instantaneous intensity becomes a random spike function. 
In the current literature statistical analysis tends to be limited to the average intensity,
 $$I={1\over T}\int_0^TdtI(t).\eqno(2.2)$$
and various types of correlation functions, the most familiar being the
auto-correlation function  
$$A(\tau)={1\over I^2}\Big[{1\over 
T}\int_0^{T-\tau}dtI(t)I(t+\tau)-I^2\Big]-\delta_\epsilon(\tau)\eqno(2.3)$$ for $T>>\tau$, 
which decays with a characteristic time scale
$\tau_D$, the diffusion time through the collection volume. The subtraction of $\delta_\epsilon$ 
permits $A(t)$ to vanish for all $\tau$ for a Poissonian histogram.
Dividing the integration time into $N_b$ bins of equal time 
interval $\tau_b$, i.e., $T=N_b\tau_b$, the number of 
photon counts falling within the time bins are 
$m_1,m_2,...,m_{N_b}$, and their moments 
$$M_l={1\over N_b}\sum_{n=1}^{N_b}m_n^l\eqno(2.4)$$ carry the 
structural information of the underlying fluorescence 
molecules.  The quantity for comparison, called the Mandel's Q 
parameter is defined as $$\delta\equiv{M_2-M_1^2-M_1\over M_1}\eqno(2.5)$$  where the 
first moment is given by 
$M_1=I\tau_b$, etc. Although higher moments yield greater 
differientiation, we limit our analysis to the second moment 
due to the experimental collection capabilities ($\leq 2^{16}$ 
photons) of our system.
The definition (2.3) and the relation $m_n=\int_{(n-1)\tau_b}^{n\tau_b}dtI(t)$ 
directly relates the Q parameter to the autocorrelation function 
$$\delta={2I\over\tau_b}\int_0^{\tau_b} ds(\tau_b-s)A(s).\eqno(2.6)$$
Similar relations exist between higher order correlation functions and higher 
order binning moments.

Finally we describe the time delay histogram of the distribution of 
time intervals between two successive photon events, i.e. $\Delta t_1$,
$\Delta t_2$,.... For sufficiently large numbers of photon counts, a distribution function 
of $\Delta t$, $\rho(\Delta t)$ can be extracted and is analogous 
to the photon waiting time distribution of quantum optics. 
\cite{Montroll75} The theory of this function will be developed 
in the sections IV and V and will be compared with experimental 
results in sections V and VI.

\section {A Mathematical Theory of Time Delay Histogram}

While there exist many articles in the literature on the mathematical properties of the photon 
event histogram \cite{Chen99,Kask99}, here we would like to provide a 
unified approach, which ties the experimental observables such as fluorescence intensity,
auto-correlation, the binning moments, and time delay histogram to a probability generating functional.

\subsection {General formulation}

To begin, 
we divide the integration time $T$ into ${\cal N}$ bins, each of 
interval $\epsilon$, i.e., $T={\cal N}\epsilon$. 
Each fluorescence diffusion process produces a histogram of photon 
counting, $\{m_1,...,m_{\cal N}\}$, with $m_l$ 
the number of photon events within the $l$-th time bin and the corresponding fluorescence intensity given by 
$$I_l={m_l\over\epsilon}\eqno(4.1)$$ Let $P_{m_1...m_{\cal N}}$ 
stand for the probability of this particular time delay histogram. The generating function of this set of 
probabilities is defined as
$${\cal G}(z_1,...,z_{\cal N})=\sum_{m_1,...,m_{\cal 
N}}P_{m_1...m_{\cal N}}z_1^{m_1}...z_{\cal N}^{m_{\cal N}},\eqno(4.2)$$ 
which is properly normalized, i.e.${\cal G}(1,...,1)=1$. In the 
limit $\epsilon\to 0$ (or, equivalently 
${\cal N}\to\infty$ with a fixed $T$), the sequences 
$\{z_1,...,z_{\cal N}\}$ and $\{I_1,...,I_{\cal N}\}$ 
become two functions of $t$, $z(t)$ and $I(t)$, such that 
$z(l\epsilon)=z_l$ and $I(l\epsilon)=I_l$. In particular, 
as $\epsilon\to 0$, most $m$'s vanish, few of them are equal to one, 
and the probability of $m_l>2$ becomes negligible. 
The function $I(t)$ approaches the random spike function introduced 
in the last section. In the same limit, 
the generating function (4.2) becomes a generating functional
$${\cal G}[z(t)]=\lim_{{\cal N}\to\infty}{\cal G}
(z_1,...,z_{\cal N}).\eqno(4.3)$$ An important set of 
observables are various correlation functions 
given by the functional derivatives of ${\cal G}$ 
with respect to $z(t)$ at $z(t)=1$, i.e. 
$$C_n(t_1,...,t_n)={\delta\over\delta z(t_1)}...{\delta\over\delta z(t_n)}\ln{\cal G}[z(t)]\mid_{z(t)=1}.\eqno(4.4)$$ 
The function $C_1(t)$ is nothing but the ensemble 
average of the instantaneous fluorescence intensity at the 
moment $t$, $$<I(t)>=C_1(t),\eqno(4.5)$$ the 
coefficient $C_2(t_1,t_2)$ is related to the auto-correlation 
function between $t_1$ and $t_2$ via 
$$A(t_1,t_2)={C_2(t_1,t_2)\over C_1(t_1)C_1(t_2)},\eqno(4.6)$$ and 
the coefficients $C_n(t_1,...,t_n)$'s 
represent higher order correlations. For the observation times, 
$t_1,...,t_{\cal N}$ sufficiently away from the 
beginning of the integration time so that transient process maybe 
ignored, these functions depend only on time differences. 
In particular, $C_1(t)$ becomes a constant and $C(t_1,t_2)$ depends only on the 
$\tau=t_2-t_1$. In this way, we can reconcile the experimentally 
defined fluorescence intensity at time $t$ and the auto-correlations of the previous section.

The distribution function of the time interval $\Delta t$ between two successive 
photon events, $\rho(\Delta t)$, referred to as $\Delta t$-distribution, can also be extracted the 
probability generating functional
$$\rho(\Delta t)={{\rm{const.}}\over T}\int_0^T d\tau{\delta\over\delta 
z(\tau)}{\delta\over\delta z(\tau+\Delta t)}{\cal G}[z(t)]
\mid_{z(t)=\zeta(t|\tau-\Delta t)},\eqno(4.7)$$ where  the characteristic function 
$\zeta(t|\tau,\Delta t)=1$ for $t\in(\tau,\tau+\Delta t)$ and $\zeta(t|\tau,\Delta t)=0$ 
otherwise, and the constant is determined by the 
condition $$\int_0^Td\Delta t\rho(\Delta t)=1.\eqno(4.8)$$ We refer the reader to the Appendix A for its derivation. 
While eqs.(4.7) and (4.8) are valid for a finite $T$, the limit $T\to\infty$ will be assumed for 
their applications. For a Poissonian histogram, (4.7) implies 
$\rho(\Delta t)=\lambda e^{-\lambda\Delta t}$, as expected.

What is relevant to the actual observation is the detected photons 
rather than the total number of emitted ones. 
Let $P_{m_1,...,m_{\cal N}}^{\rm{eff.}}$ denoted the 
probability of a particular histogram $\{m_1,...,m_{\cal N}\}$ of detected photons. 
The corresponding generating function reads $${\cal 
G}_{\rm{eff.}}(z_1,...,z_{\cal N})=\sum_{m_1,...,m_{\cal N}}P_{m_1m_2...m_{\cal N}}^{\rm{eff.}}z^{m_1}...z^{m_{\cal N}},\eqno(4.9)$$ As is shown in the appendix A, under 
the assumption that all photon counts are statistically independent, we 
find a simple relation
$${\cal G}_{\rm{eff.}}(z_1,...,z_{\cal N})={\cal G}(1-\eta+\eta z_1,...,1-\eta+\eta z_{\cal N})\eqno(4.10)$$ 
with $\eta$ the efficiency of the detector. In the limit ${\cal N}\to\infty$, 
it becomes $${\cal G}_{\rm{eff.}}[z(t)]={\cal G}[1-\eta+\eta 
z(t)].\eqno(4.11)$$
 
\subsection{A Quantum Mechanical Analog}
The formulation we have established thus far is completely general, 
and independent of details of the fluorescence 
diffusion process. We shall now include the details of 
the physical process and model the generating functional ${\cal G}[z(t)]$.  
Consider $N$ molecules, each having a specific brightness, and 
diffusing in a solution of total volume $\Omega$. Both 
$\Omega\to\infty$ and $N\to\infty$, at a fixed 
concentration $c={N\over\Omega}$. An axially 
symmetric intensity profile is created by focusing the laser beam within the sample solution. While fluorescence occurs everywhere along 
the beam volume, the pinhole effectively eliminates 
the collection of photons emitted far away from the focal point 
\cite{Rigler93} 
and a small detection volume is defined, which contains few molecules in average 
at all times. In the absence of chemical 
reactions (no self hybridization), sufficiently weak laser intensity and 
sufficiently low concentrations, we may assume that i) 
$N$ molecules do not interact with each other; ii) photon emissions are 
statistically independent in the absence of diffusion; and 
iii) the photon emission frequency per fluorophore within the 
detection volume is $\lambda_0u(\vec r)$.  $u(\vec r)$ will be
referred to as the fluorescence profile function and is normalized 
by the condition $u(0)\equiv u_{\rm{Max.}}(\vec r)=1$.  Under these assumptions, 
the probability generating functional takes the same mathematical expression of the 
transition matrix element of a quantum mechanical system of $N$ noninteracting bosons 
in an external 
potential field and imaginary time, i.e.  $${\cal G}[z(t)]=\Big({1\over\Omega}\Big)^{\cal N}<|{\cal T}
e^{-\int_0^TdtH(t)}|>,\eqno(4.12)$$ where ${\cal T}$ enforces the time ordering and 
$\Omega$ is the volume of the solution and will be sent to infinity for all practical purposes. 
The Hamiltonian operator of the analog quantum mechanical system reads 
$$H(t)=\sum_{j=1}^Nh(\vec r_j,t)\eqno(4.13)$$ and
$$h(\vec r,t)=-D\nabla^2+[1-z(t)] \lambda u(\vec r),\eqno(4.14)$$ 
where $D$ is the 
diffusion constant and $\lambda=k\lambda_0$ with $k$ the effective number of fluorophores per molecule. 
The wave function of the state $|>$ of the analog quantum 
mechanical system is of zero momentum and is normalized in the coordinate representation to 
$$<\vec r_{1},...,\vec r_{\cal N}|>=1.\eqno(4.15)$$
The full derivation of (4.12)-(4.14) are presented in the appendix B. 
We notice that:\\

1). The Hamiltonian operator (4.14) describes the motion of a particle of mass $(2D)^{-1}$ moving 
in an external potential $[1-z(t)]\lambda u(\vec r)$. Two kinds of expansions can be developed 
for the statistical analysis. The first is a perturbative expansion according to
the powers of $1-z(t)$, which generates the correlation functions to all orders. The 
second is the expansion according to the powers of the diffusion constant, $D$, which is 
particularly useful for FCS with biological molecules. The leading order of the second 
expansion corresponds to the frozen limit in the literature 
\cite{Chen99,Kask99} and we are able to add the higher order corrections 
systematically following this quantum mechanical 
analog.\\
  
2). It follows from (4.11) and (4.14) that the generating functional responsible 
for the observed time delay 
histogram, ${\cal G}_{\rm{eff.}}[z(t)]$, assumes the identical mathematical 
form as  ${\cal G}[z(t)]$, provided 
$\lambda$ is replaced by $\lambda_{\rm{eff.}}=\eta\lambda$. In what follows, we shall refer exclusively to 
${\cal G}_{\rm{eff.}}[z(t)]$ with the subscript ``eff.'' suppressed.\\

3). The generating functional (4.12) can be factorized for each 
aggregate, i.e.
$${\cal G}[z(t)]=g^N[z(t)]\eqno(4.16)$$
with
$$g[z(t)]={1\over\Omega}<|{\cal T}e^{-\int_0^Tdth(\vec r,t)}|>\eqno(4.17)$$
and $<\vec r|>=1$. Alternatively, $g[z(t)]$ can be calculated by integrating the wave function $\psi(\vec r,t)$ 
that solves the Schroedinger equation with an imaginary time,
$${\partial\over\partial t}\psi(\vec r,t)=-h(\vec r,t)\psi(\vec r,t)\eqno(4.18)$$
subject to the initial condition, $\psi(\vec r,0)=1$, i.e.
$$g[z(t)]={1\over\Omega}\int d^3\vec r\psi(\vec r,T).\eqno(4.19)$$
The differential equation (4.18) and its initial condition below can be converted to an integral equation 
by treating the potential term of $h(\vec r,t)$ as a source of the diffusion,
$$
\psi(\vec r,t)=1+{\lambda\over(4\pi D)^{3\over 2}}\int_0^tdt^\prime{1-z(t^\prime)
\over(t-t^\prime)^{3\over 2}}\int d^3\vec r^\prime e^{-{(\vec r-\vec r^\prime)^2\over 4D(t-t^\prime)}}
u(\vec r^\prime)\psi(\vec r^\prime,t^\prime).\eqno(4.20) 
$$
It follows from (4.19) that
$$g[z(t)]=1+{\lambda\over\Omega}\int_0^tdt^\prime[1-z(t^\prime)]\int d^3\vec r^\prime u(\vec r^\prime)
\psi(\vec r^\prime, t^\prime).\eqno(4.21)
$$
As the main contribution to the integral comes from the detection volume specified by $u(\vec r)$, we find 
$$
g[z(t)]-1=O(1/\Omega)\eqno(4.22)$$
in the limit $\Omega\to\infty$ with a fixed detection volume. Taking this limit at a fixed concentration,
$c$, and using the standard limit, $\lim_{N\to\infty}(1+{x\over N})^N=e^x$,
we obtain that
$${\cal G}[z(t)]=e^{-{\cal F}[z(t)]}\eqno(4.23)$$
with 
$${\cal F}[z(t)]=c<|1-{\cal T}e^{-\int_0^Tdth(\vec r,t)}|>.\eqno(4.24)$$\\

\subsection{Generalization to Multi-species and Multi-channels}
We generalize the present formulation to include several species of 
fluorescent molecules with multiple channels of detection. Assuming there 
are $M$ species each labeled 
by an index $l$ and $K$ detecting channels each labeled by $\alpha$, the Hamiltonian of the 
analog quantum mechnical system, (4.13), becomes
$$H(t)=\sum_{l=1}^M\sum_{j=1}^{N_l}h_l(\vec r_j,t)\eqno(4.25)$$
with $$h_l(\vec 
r,t)=-D_l\nabla^2+\sum_{\alpha=1}^K[1-z_\alpha(t)]\lambda_{l\alpha}(\vec r)\eqno(4.26)$$
where $\lambda_{l\alpha}(\vec r)$ the profile function specific to the $l$th species and $\alpha$th 
channel, and it becomes $\lambda u(\vec r)$ for a single species and a single channel. The generating 
functional (4.12) factorizes into a product of a single species, where each
now depends on several arbitrary functions $z_\alpha(t)$. The power series expansion according to 
$1-z_\alpha(t)$'s yields all the corresponding correlation functions. Unlike 
number counting in the frozen limit, it can be factorized 
into individual detecting channels for nonzero diffusion constants.

\section {Analytical Expressions for Data Analysis}
In this section, we shall display the analytical expressions for fluorescence intensity, 
the auto-correlation function, Mandel's Q parameter and the $\Delta t$ distribution, as derived from the general 
formulation of the previous section. The technical details of the derivation are deferred to the Appendix C. 

\subsection{Fluorescence intensity, correlation functions and the binning statistics}
\noindent In accordance with eqns.(4.4) and (4.5), the ensemble average of the fluorescence intensity reads
$$I=c\lambda\int d^3\vec ru(\vec r)=n\lambda,\eqno(5.1)$$ where $n=cv$ with $v\equiv\int d^3\vec ru(\vec r)$. 
The integration $v$ can be viewed as the effective collection volume 
defined by the focused beam and pinhole and $n$, the average number 
of molecules within the volume.  The experiments reported in this article 
are characterized by $n\sim 1$.\\
  
\noindent Applying the definition (4.4) and the quantum mechanical analog 
(4.12)-(4.14) of ${\cal G}[z(t)]$, the second 
order correlation $C_2(t_1,t_2)$ takes the form $$C_2(t_1,t_2)=c\lambda^2\int{d^3\vec p\over (2\pi)^3}u_{\vec 
p}^2 e^{-|t_1-t_2|Dp^2}\eqno(5.2)$$ with $u_{\vec p}=\int d^3\vec re^{-i\vec p\cdot \vec r}u(\vec r)$ the 
Fourier transformation of the profile function $u(\vec r)$. Substituting 
(5.1) and (5.2) into (4.6) for the 
auto-correlation function, we find:\\

\noindent 1). For an arbitrary $u(\vec r)$, the auto-correlation function at zero time 
lag takes a simple form $$A(0)={Z\over n}\eqno(5.3)$$ revealing the 
number or concentration of objects within the volume and $Z$ the geometrical 
factor for that volume defined by $$Z={\int d^3\vec ru^2(\vec r)\over \int d^3\vec ru(\vec 
r)}.\eqno(5.4)$$ For nonzero time lag, we shall parametrize the auto-correlation function as 
$$A(\tau)={Z\over n}{\cal A}(\tau)\eqno(5.5)$$ with ${\cal 
A}(0)=1$.\\  
  
\noindent 2). For a 3D Gaussian fluorescence profile, $$u(\vec 
r)=e^{-{x^2+y^2\over 2\omega_\perp^2}-{z^2\over 2\omega_{\|}^2}},\eqno(5.6)$$
we obtain that $Z={1\over 2\sqrt{2}}$ and [1] $${\cal A}(\tau)={1\over(1+{\tau\over\tau_\perp})\sqrt{1+{\tau\over\tau_\|}}}\eqno(5.7)$$
with $\tau_\perp=\omega_\perp^2/D$ and $\tau_\|=\omega_\|^2/D$.  For a Gaussian-Lorentzian profile, 
$$u(\vec r)={\omega_\|^2\over \omega_\|^2+z^2} e^{-{x^2+y^2\over2\omega_\perp^2}},\eqno(5.8)$$ 
we find that $Z={1\over 4}$ and $${\cal A(\tau)}={2\over(1+{\tau\over\tau_\perp})}\sqrt{{\tau_\|\over\tau}}e^{\tau\over\tau_\|}{\rm{Erfc}}\Big(\sqrt{\tau_\|\over\tau}\Big) \eqno(5.9)$$ with 
$${\rm{Erfc}}(z)=\int_z^\infty dx e^{-x^2}={e^{z^2}\over 2z}\Big[1-{1\over 
2z^2}+O(z^{-4})\Big].\eqno(5.10)$$ 
For $\tau<<\tau_\|$ and $\tau_\perp<<\tau_\|$, both (5.7) and (5.9) 
can be approximated by $${\cal A}(\tau)={1\over 
1+{\tau\over\tau_\perp}}.\eqno(5.11)$$ 
Extending the same analysis to the third order coefficient of the expansion of $\ln{\cal G}$ according to the power of $z(t)-1$, we find the third order 
correlation function for $\tau_\perp<<\tau_\|$, i.e.
$$C_3(t_1,t_2,t_3)={Z^\prime\over(1+{\tau\over\tau_\perp})(1+{\tau^\prime\over\tau_\perp})-{1\over 4}},\eqno(5.12)$$ 
where the constant $Z^\prime$ is another geometrical factor, like $Z$ for 
the autocorrelation, $\tau\equiv t_a-t_b$, $\tau^\prime\equiv t_b-t_c$ with $t_a$, $t_b$ and $t_c$ a 
permutation of $t_1$, $t_2$ and $t_3$ such that $t_a>t_b>t_c$.

Using Eqn(4.6), we obtain the expression of the Q parameter 
which agrees with that of FIMDA \cite{2Kask00}. 
For $\tau<<\tau_\|$ it can be approximated by
$$\delta = 2Z\lambda\tau_\perp\Big({\tau_b+\tau_\perp\over\tau_b}\ln{\tau_b+\tau_\perp\over\tau_\perp}-1\Big).\eqno(5.13)$$
with $\tau_b$ the size of the binning window and the dependence on different models of longitudinal profile absorbed in the constant $Z$.
The distribution of photon counting numbers within a time bin can be extracted from the generating function $G(z)$, obtained from the generating 
functional ${\cal G}[z(t)]$ by restricting the form of $z(t)$ such that it equals to a constant $z$ withing the time bin and vanishes elsewhere.
We then have 
$$G(z)=e^{-nF(\tau|1-z)}\eqno(5.14)$$
with 
$$F(\tau|\zeta)\equiv {c\over n}<|1-e^{-\tau h(\zeta)}|>\eqno(5.15)$$
and
$$h(\zeta)=-D\vec\nabla^2+\zeta\lambda u(\vec r).\eqno(5.16)$$
While an analytical expression for $F(\tau|\zeta)$ does not exist in general, the expansion according to diffusion 
constant can be obtained easily, 
$$F(\tau|\zeta)=\zeta\lambda\tau f(\zeta\lambda\tau)-{1\over 
3\lambda\tau_\perp}\zeta^2\lambda^2\tau^3f^\prime(\zeta\lambda\tau)+O({1\over\lambda^2\tau_\perp^2})\eqno(5.17)$$
with $$f(x)={4\over 3\sqrt{\pi}}\int_0^1d\xi\ln^{3\over 
2}{1\over\xi}e^{-x\xi},\eqno(5.18)$$
where the leading term corresponds to the frozen limit in \cite{2Kask00} and the second term improves the approximation further.
  
\subsection{The time-delay histogram}
  
\noindent Carrying out the functional derivatives in the formulae (4.7) 
with the the aid of (4.16) and (4.17), the distribution function 
$\rho(\Delta t)$ becomes $$\rho(\Delta t)={1\over\lambda n}{d^2\over d\Delta 
t^2}e^{-nF(\Delta t|1)}=\lambda\Big[-{\partial^2F\over\partial(\lambda\Delta t)^2}
+n({\partial F\over\partial(\lambda\Delta t)})^2\Big]e^{-nF(\Delta t|1)} \eqno(5.19)$$ 
Using expansion (5.17), we derive an approximate expression of $\rho(\Delta t)$ 
which is valid for $\lambda\tau_\perp>>1$ and
$\Delta t<<\tau_\perp$,
$$\ln\rho(\Delta t)=\ln\rho(0)-nxf(x)
+\ln\lbrace[f(x)+xf^\prime(x)]^2-{1\over n}[2f^\prime(x)+xf^{\prime\prime}(x)]\rbrace$$
$$+{1\over\lambda\tau_\perp}\lbrace-{1\over 3}nx^3f^\prime(x)
+{2xf^\prime(x)+2x^2f^{\prime\prime}(x)+{1\over 3}x^3
f^{\prime\prime\prime}(x)-2nx^3[f(x)+xf^\prime(x)][f^\prime(x)-{1\over 3}xf^{\prime\prime}(x)]\over
-2f^\prime(x)-xf^{\prime\prime}(x)+n[f(x)+xf^\prime(x)]^2}\rbrace\eqno(5.20)$$
with $x=\lambda\Delta t$. Alternatively, a Taylor expansion of $F$ in 
$\lambda\Delta t$ yields an expression of $\rho(\Delta t)$ for 
$\lambda\Delta t<<1$, which applies to the case with arbitrary 
$\lambda\tau_\perp$.

\noindent For Taylor expansion of the function $F$ according to the power of $\Delta t$, 
we obtain $$\ln\rho(\Delta t)=\ln\rho(0)-(n+\gamma)\lambda\Delta 
t+{1\over 2}(\beta-\gamma^2)\lambda^2\Delta t^2+O(\Delta t^3),\eqno(5.21)$$ 
where $$\gamma={3^{-{3\over 2}}+2^{-{1\over 2}}n+{2^{-{3\over 2}}\over\lambda\tau_\perp}\over 2^{-{3\over 2}}+n}$$
and $$\beta= (2^{-{3\over 2}}+n)^{-1}\Big[{1\over 8}+({3\over 8}+2\cdot3^{-{3\over 2}})n+2^{-{3\over 2}}n^2
+{4\cdot3^{-{5\over 2}}+2^{-{1\over 2}}n\over\lambda\tau_\perp}\Big].\eqno(5.22)$$
If the histogram were a Poissonian, a single exponential would be expected, 
which corresponds to $\beta=\gamma=0$. The parameters $\beta$ and $\gamma$ represent the deviation from 
a Poissonian which do not vanish even in the frozen limit, i. e. $\tau_\perp\to\infty$. At high density, on the 
other hand, $n>>1$, $\beta=O(n)$ and $\gamma=O(1)$, eq.(5.21) can be written as 
$$\ln\rho(\Delta t)=\ln\rho(0)-z+O({z^2\over n}),\eqno(5.23)$$
with $z=n\lambda\Delta t$which agrees with that of a single Poissonian. 

\section{Experimental Materials and Methods}
  
To create well-defined single or double fluor elements, short pieces 
of single stranded DNA (ssDNA) 
were used as substrates.  Either one or two fluorophores can be 
site-specifically coupled to each DNA oligomer, (29 
bases in length) dependent on the number of end-strand modifications 
(primary amino linker arms, 
Midland Certified Reagent Co.). Succidinmyl ester Rhodamine 6G 
(Molecular Probes) was coupled to the 
modified sites in the presence of DMF, and purified by gel filtration 
and reverse phases chromatography 
(HPLC).  

\begin{figure}
\centerline{\epsfxsize 6in \epsfbox{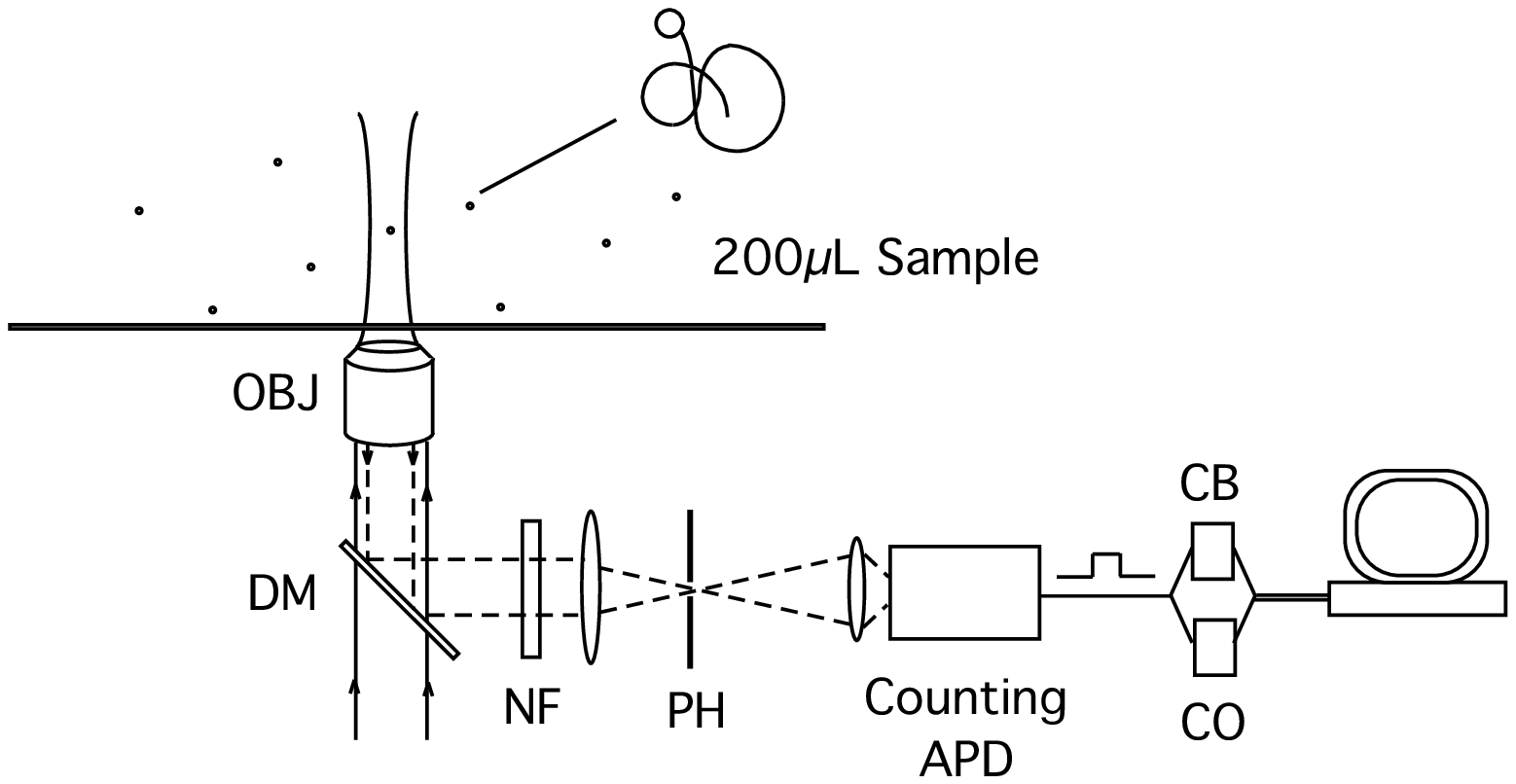}}
\vskip0.4cm
\caption{}
\label{fig:setup}
\end{figure}

The experimental setup (Fig.\ref{fig:setup}) is an inverted confocal 
microscopy arrangement.  The sample is illuminated by 
the 514.5nm line of a Ar+ laser (Lexel 85) focused through a 60x 
water immersion objective 
(numerical aperature 1.2, Olympus).  Incident power was empirically 
optimized 
at 100$\mu$W, so that
the photon counts per aggregate were at least 50,000 cps, while 
avoiding significant population of the 
triplet state or bleaching.  Emitted photons are collected through 
the same objective, directed through 
a high-pass dichroic mirror (Omega Optical) and a notch filter 
(Kaiser Optical) to reduce collection of 
on-axis elastically scattered photons. The collection volume is 
further refined by focusing the light 
onto a 25$\mu$m pinhole, eliminating off angle scattering as well as 
spatially defining the collection volume.  The collection volume was 
empirically determined though the number of molecules measured through FCS as a function 
of increasing concentration.  All FCS measurements were 10 minutes in 
duration using a ALV 5000 E board for data collection and in-house 
data analysis software for fitting.    
The overall detection efficiency of the setup is estimated to be ~3 percent.  Photons 
are detected by a counting avalanche photodiode (EG\&G/Perkin Elmer), 
pulsewidth ~25ns 
whose signal is processed by task specific counting board (NI-TIO 
6602 National Instruments), 
controlled by LabVIEW.  The period between each photon is measured by 
counting the number of 
external clock (4MHz) pulses(source) between each photon pulse 
(gate).  $2^{16}$ photons are collected for 
each trial.  Data acquisition technology limited the total collection 
time to 1-2s, depending on the incident intensity. Afterpulsing artifacts from the photodiodes were 
measured $\sim$ 100ns, with comparison to cross-correlation of the same 
signal in two perpendicular detectors.  Digital filtering was used to 
subtract 
the afterpulsing noise from the final histogram statistics.  DNA 
concentrations 
are $\sim$nM in PBS buffer such that only one monomer molecule at any time is 
within 
the collection volume.

Photon counting data is collected as fluorescent 
particles diffuse through the sample volume.
When the particles are outside the volume they are 
dark and undetectable.  As they enter the volume they are excited 
with a certain probability and 
emit a temporal pattern of photons that are detected by a counter.  Given a 
long integration time, many particles diffuse through the volume 
producing temporal fluctuations in fluorescence.   Whereas the 
autocorrelation function reveals 
the timescale of these fluorescence fluctuations, the 
probability distribution characterizes their amplitude.

\section{Results}

\begin{figure}
\centerline{\epsfxsize 6in \epsfbox{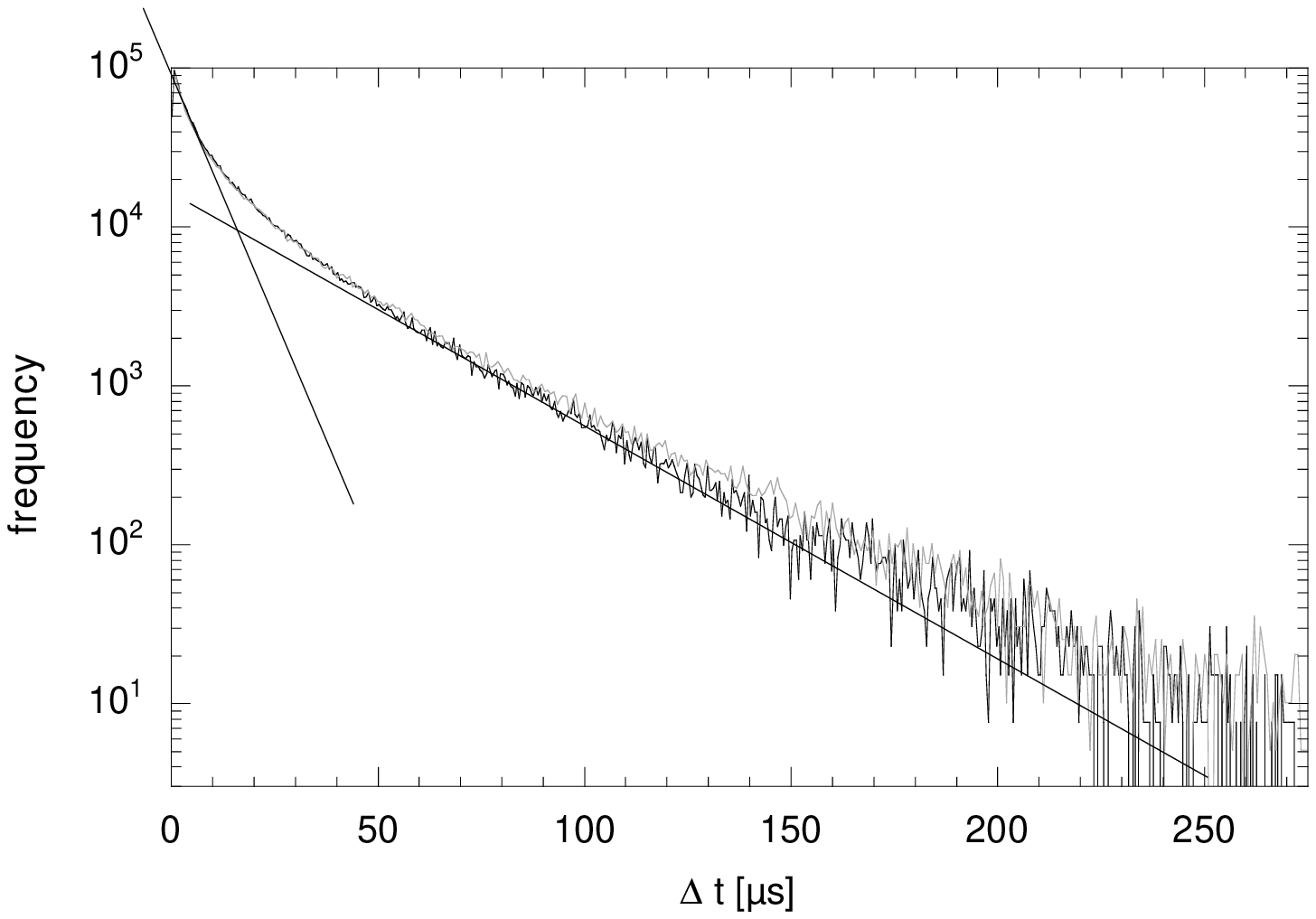}}
\vskip0.4cm
\caption{}
\label{fig:comp}
\end{figure}

Photon counting data was collected for two systems, and the time 
delay histogram constructed for each (Fig \ref{fig:comp}).  Both 
samples consist of very dilute (nM) identical sequences of ssDNA in 
buffer, and observations 
were made at room temperature (~25 \degc).  We consider these systems 
noninteracting. 
In the first sample, specifically single end-labeled ssDNA  
was diluted to an average concentration of  1 
molecule per collection beam volume at any time.   The number of molecules was calculated from 
calibration of the collection volume (see Methods) at $100\mu W$ incident power (0.35$\mu m^{3}$). 
The second sample contained the 
same sequence ssDNA however tagged with two fluorophores 
per object (one at each end).  This sample was then diluted so that the average 
intensity in time 
of the two samples were {\em equal}.  If the dimer system was exactly 
twice the fluorescence of the monomer, then the dimer concentration 
should be 0.5 that of the monomer for the same given intensity.  However, 
due to local quenching effects of 
the fluorophore, the average concentration of double-dyed molecules 
in the beam at any time was calculated to be 0.7 to maintain the same intensity.
The empirical volume of the beam was 
calibrated from a serial dilution of a standard dye solution, and the
rate of emitted photons per fluor,$\lambda$, measured. $1/\lambda = 1/60kHz$.  Concentrations 
of single/double fluorescent aggregates in the beam were 1 and 0.7 
molecules respectively.    

Empirical measurements demonstrated the quenching could be minimized  
by hybridization of the ssDNA to a non-fluorescent target.  Hybridization of such a 
short segment (below the persistent length of dsDNA) forces the two fluorophores 
farther apart, minimizing dye-dye interaction.  However, double-dyed molecules never 
demonstrated a full factor of 2 increase 
in fluorescence over their single-dyed conterparts.  We suspect local 
quenching interaction of the fluor with the nearby 
base of the target strand.

The characteristic 
diffusion time through the sampling volume 
was extracted from the decay of the autocorrelation function.  Due to 
the resolution of our correlator software $(10^{-8}s-10^{-2}s)$, we 
are most 
sensitive to the  diffusion across the short axis of the beam 
profile.  Hence, the characteristic time extracted from time 
correlation of the data is most representative of 
$\tau_\perp$. Through FCS measurements, we find that 
$\tau_\perp\simeq 300\mu$s for both single-dyed molecules and double-dyed molecules, 
Using a random walk simulation and a geometical approximation 
for the long axis of the beam profile, we estimate $\tau_\|$ to be 
$100\tau_\bot$.  

In all theoretical curves of the figures below, we use $\lambda=60$kHz for specific brightness of 
the single-dyed molecules and an enhancement of $1/0.7\simeq 
1.4\lambda$ for the double-dyed molecules, 
The transverse diffusion time $\tau_\perp=300\mu$ s is substituted into the theoretical formula 
for for both single-dyed and double-dyed molecules. 

\begin{figure}
\centerline{\epsfxsize 6.5in \epsfbox{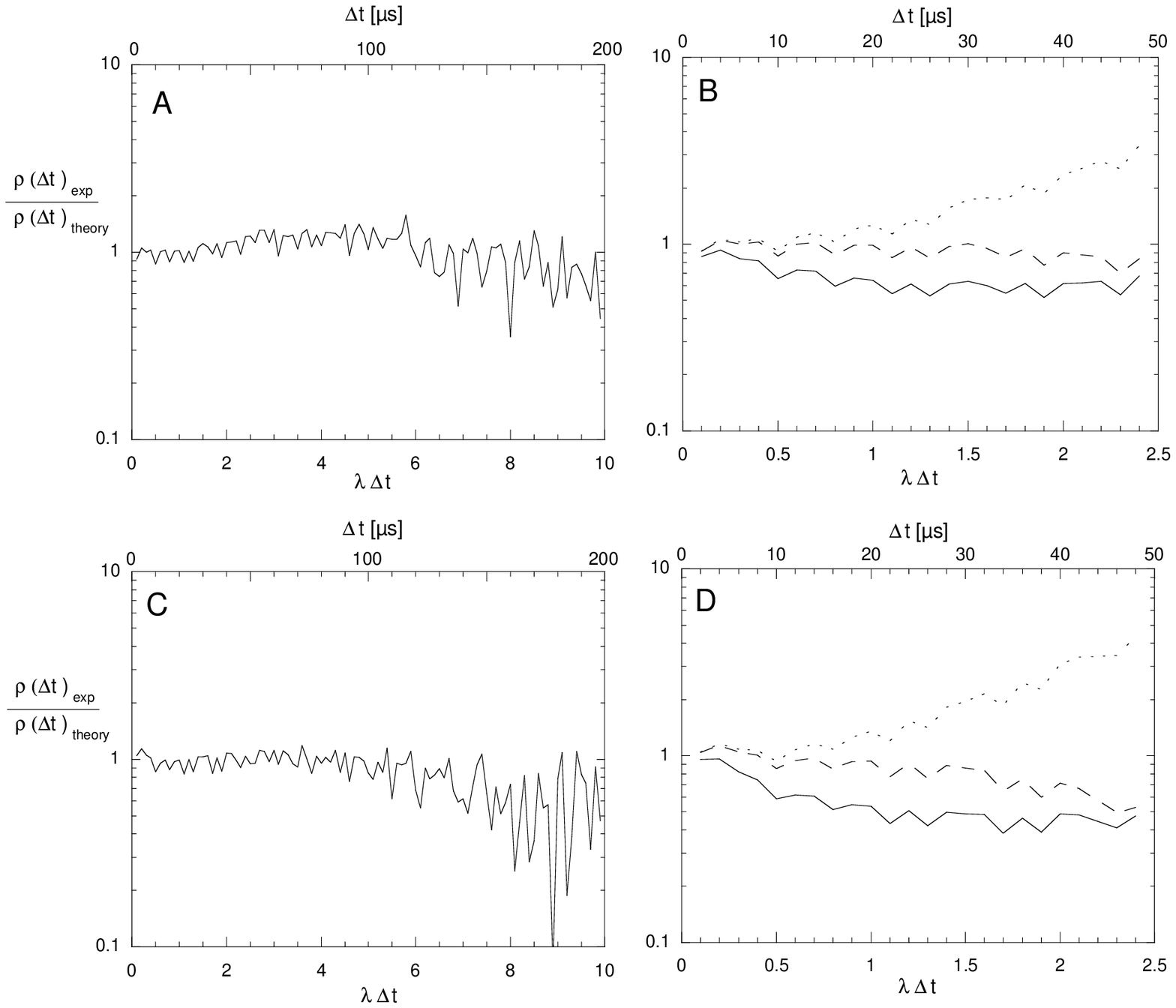}}
\vskip0.4cm
\caption{}
\label{fig:fit}
\end{figure}

At first glance, the two distributions of Fig.3 are completely indistinguishable.  
However upon closer examination it is possible to see the two curves 
slowly diverge at large $\Delta t$ with the double dye data falls 
slightly above the single dye data. The comparison between the analytical expression of the 
time delay histograms with large and small $\Delta t$ approximations (5.20) and (5.21) 
is shown in Fig.\ref{fig:fit}, where the ratio of the experimental histogram to the theoretical 
one is plotted versus the time delay $\Delta t$.   
It is evident that the formula (5.19) together with the approximation (5.17), 
denoted as $\rho_{\rm{theory}}(\Delta t)$ in the Figures, is  robust for single (A) and 
double dyed (C)  samples.  (B) and (D) are exploded 
views of the small $\Delta t$ domain of each histogram. The curve labeled by ``linear'' or 
``quadratic'' corresponds to the theoretical formula (5.21) with terms beyond linear or 
quadratic truncated. The quality of the agreement is improved from the linear truncation 
to the quadratic truncation. The curve labeled by ``Poisson'' corresponds to 
$\rho_{\rm theor.}(\Delta t)$ given by a Poissonian, i.e. $\ln\rho_{\rm theor.}
(\Delta t)=\ln\rho_{\rm theor.}(0)-\lambda\Delta t$, which is clearly a poor 
description of the experimental histogram.

Although these simulations  successfully differentiate the two 
systems, the analysis is somewhat cumbersome.  A common mathematical 
technique to highlight the subtle differences in 
distributions is moment analysis.  Similar techniques have 
proven useful in fluctuation spectroscopy
\cite{Palmer87,Qian90,2Qian90}.  The first moment is the 
mean, the second moment the standard deviation, the third the 
skewness, etc.  

Returning to Fig.3, we note a profound 
difference between the experimental data and a single Poissonian 
process.  The straight lines represent two hypothetical single Poission 
systems with different timescales.  For the simple detection of emitted photons 
within a fixed volume one might expect the statistics to resemble a single 
Poissonian process. \cite{Mandel58}  However, when the particles are allowed to 
diffuse through the boundaries of the volume, an additional Poissonian process 
contributes to the overall photon statistics. \cite{Snyder75,Mandel95} Not only 
must we account for the  stochastic nature of the emission process, we must 
also consider the number distribution of aggregates passing into the beam 
volume from the larger sample reservoir.  The statistics of the time series of 
the photon counting can be highlighted through Mandel's Q 
parameter (Eqn 2.5) introduced in section II.  

\begin{figure}
\centerline{\epsfxsize 6in \epsfbox{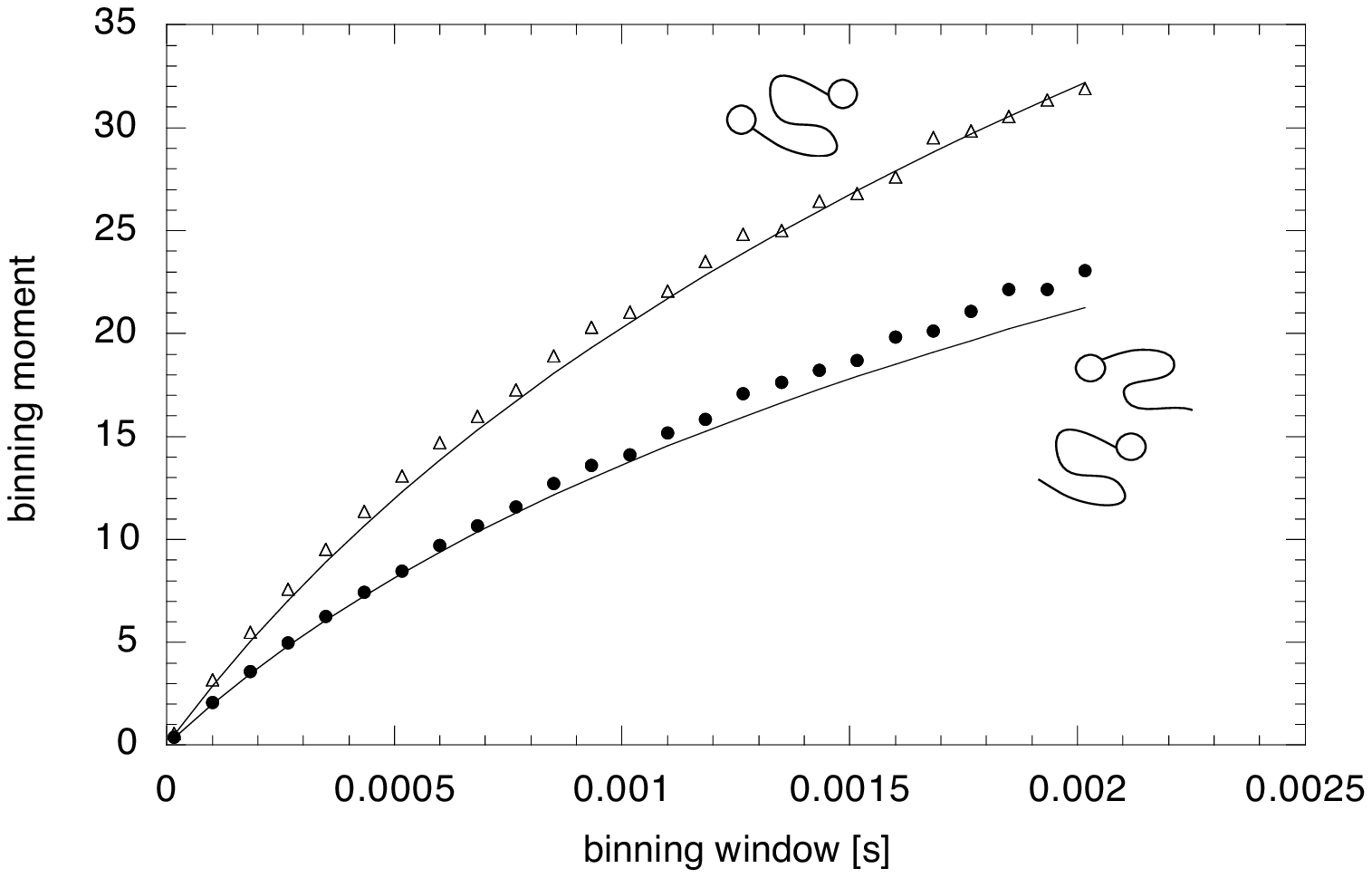}}
\vskip0.4cm
\caption{}
\label{fig:binmom}
\end{figure}

We develop the binning moment as a complementary technique to FIMDA.  In 
FIMDA \cite{Kask00} 
each histogram is representative of the number count distribution using a 
certain binning window size.  For every change in binning window size, a new 
histogram is constructed.  Likewise, every histogram has its own unique set of 
moments. All first moments should be equal to the product of the average intensity 
and the the window size and will 
not show any difference between the two systems (single dye and double dye) in accordance with 
the experimental procedure described in the first paragraph of this section. 
Starting with the 
second moment, the difference between the two system emerges. In Fig \ref{fig:binmom}, we plot 
the second moment normalized according to (2.5) for both system and the corresponding 
theoretical curve given by (5.13). The data for the two systems are clearly distinguishable and 
agree well with the theoretical prediction.

If the photon histogram were a single Poissonian, all correlations as well 
as the Q parameter would vanish. Therefore the correlation 
functions and the Q parameter measure the deviation of the photon histogram 
from a Poissonian.  Recall that the time-delay histogram is 
essentially the result of a two Poissonian processes.  If we re-examine 
Fig \ref{fig:comp} (although it is a time delay, not number counting 
histogram) and focus on the dashed
line through the short $\Delta t$ domain, divergence from the single 
Poisson increases as $\Delta t$ increases.  For binning windows smaller than 
the diffusion time, the statistics are primarily due to 
the intrinsic fluorescence of the fluorophore.  Since both systems 
contain 2 fluors, the two distributions vary little in this domain.  
At longer times, each molecule samples the inhomogeneous beam profile as it 
diffuses.  Hence the diffusion dependent statistics dominate the long 
time domain and are responsible for the notably different Q parameters of the samples.

\section{Secondary Effects}
In traditional fluorescence correlation spectroscopy, triplet state 
effects and various quenching processes are the principal
mechanisms that overshadow structural information. To
parse out these contributions, one may need to explore the higher 
binning
moments than have been examined in this work. Our mathematical 
formulation
provides the complete systematics for this purpose.  Such secondary
effects introduce additional timescales to the problem (not simply 
the 
fluorescence rate and the diffusion time to cross the collection
volume discussed in this paper). Also, some details of the electronic 
transition
inside a fluorophore should be addressed. This modeling can be 
achieved by
enlarging the quantum mechanical analog with a multi-component wave 
function $\psi_i$ 
and a matrix Hamiltonian
$$h_{ij}=-D\nabla^2\delta_{ij}+V_{ij}\eqno(7.1)$$ where
$$V_{ij}=\cases{z(t)W_{ij}&if $i\to j$ is a fluorescence transition;\cr
W_{ij}&if $i\to j$ is not a fluorescence transition;\cr 
-\sum_kW_{ik}&if $i=j$.\cr}\eqno(7.2)$$ 
Each component of the wave function, $\psi_i$ represents the
probability of a DNA molecule at a particular spatial location with 
its fluorophore in the $i$-th electronic level and $W_{ij}$ the transition 
rates from $i$th $\to$ $j$th of electronic levels.

In principle, such an elaboration should also be implemented in the 
absence of
triplet state effects and quenching processes, since the fluorescence 
rate
combines the excitation rate from the ground state and the spontaneous
emission rate from the excitation levels. For two electronic levels with '0' labeling the 
ground state and '1' the excitation state, we have $W_{01}=\lambda u(\vec r)$ and $W_{10}=\lambda_s$, 
Einstein's $A$-coefficient. The Schroedinger equation (4.18) is split into two 
components:
$${\partial\over\partial t}\psi_0=(D\nabla^2-\lambda u)\psi_0+z(t)\lambda_s\psi_1\eqno(7.3)$$
$${\partial\over\partial t}\psi_1=(D\nabla^2+\lambda_s u)\psi_1-\lambda u\psi_0\eqno(7.4)$$

The approximation employed in previous sections amounts to $\lambda_s>>\lambda$ and $\lambda>>1/\tau_\perp$, the 
diffusion time. In this case, the spontaneous emission is almost instantaneous once the fluorophore 
is excited and the second equation, (7.4), gives rise to $\lambda_s\psi_1\simeq\lambda u\psi_0$
at equilibrium. Substituting it back (7.3), we obtain (4.18) with $h$ given by (4.14). 
For the experimental data presented in this paper, a
60kHz photon time delay histogram at 3\% detection
efficiency, the excitation rate is 60/3\%=2MHz, and the
corresponding fluorescence time is 500ns, much longer than the typical
spontaneous emission time, 10ns. Our approximation is therefore 
adequate.

\section{Concluding Remarks}

We have developed a mathematical
formulation to analyze the time delay series of fluorescence photons from 
diffusing particles, 
based on a probability generating functional and its quantum 
mechanical 
analog. Although it may appear formal, since some analytical 
expressions such as the
auto-correlation function have been obtained by less sophisticated 
means, the
approach is systematic. The potential of this general approach
will be realized when dealing with systems of greater complexity, e.g. in the presence of 
the chemical reaction discussed below.

We have designed an experiment to differentiate fluorescent 
aggregates 
in solution. ssDNA monomers are 
labeled with a single fluorophore and dimers with two fluorophores.  
Secondary effects such as 
chemical reactions, triplet state effects and various quenching 
processes 
have been neglected in our model.  However 
these effects have been minimized by using dilute solutions to avoid 
self-interaction and Rhodamine 6G, a fluorophore 
with little triplet state at low incident intensities.

Although we have only addressed ideal experimental conditions of the 
non-reacting case in this manuscript, typical biological/chemical 
systems 
react (aggregation or cooperative binding). Our technique can be 
modified to include these reactions. One must generalize
the quantum mechanical analog to the case with several species of 
particles,
each representing a fluorescence molecule, interacting with each 
other. Without going to technical details which will be reported elsewhere, we quote 
the generalization of our formulation analog in the presence of a binary reaction, 
$$2A\iff B.\eqno(8.1)$$ The Hamiltonian of the quantum mechanical analog in (4.12) is given by 
$$H=H_{\rm{diff.}}+H_{\rm fluor.}+H_{\rm{chem.}},\eqno(8.2)$$
where
$$H_{\rm{diff.}}=\int d^3r(D_A\vec\nabla\bar\psi\cdot\vec\nabla\psi
+D_B\vec\nabla\bar\phi\cdot\vec\nabla\phi),\eqno(8.3)$$
$$H_{\rm fluor.}=[1-z(t)]\int d^3ru(\vec r)[
(\lambda_A\bar\psi\psi+\lambda_B\bar\phi\phi)\eqno(8.4)$$ and
$$H_{\rm{chem.}}=-{1\over 2}\int d^3r\int d^3r^\prime\int d^3\vec R
\sigma(\vec R|\vec r,\vec r^\prime)\bar\phi(\vec R)\psi(\vec r)
\psi(\vec r^\prime)$$
$$-{1\over 2}\int d^3r\int d^3r^\prime\int d^3\vec R
\sigma^\prime(\vec r,\vec r^\prime|\vec R)\bar\psi(\vec r)
\bar\psi(\vec r^\prime)\phi(\vec R)$$
$$+{1\over 2}\int d^3r\int d^3r^\prime\int d^3\vec R
\sigma(\vec R|\vec r,\vec r^\prime)\bar\psi(\vec r^\prime)
\bar\psi(\vec r)\psi(\vec r)\psi(\vec r^\prime)$$
$$+{1\over 2}\int d^3r\int d^3r^\prime\int d^3\vec R
\sigma^\prime(\vec r,\vec r^\prime|\vec R)\bar\phi(\vec R)
\phi(\vec R),\eqno(8.5)$$
and the generating functional is no longer factorizable.
In this Halmitonian, the pairs of operators ($\psi$,$\bar\psi$) or 
($\phi$,$\bar\phi$) are the creation/annihilation operators of a molecule of 
species A or B. $D_{A(B)}$ and $\lambda_{A(B)}$ are the diffusion constant and the specific brightness of the species A(B). The first term of the reaction part, 
$H_{\rm chem}$ represents the creation of a B-molecule, the second term 
represents the creation of a pair of A-molecules, the third term signifies the annihilation of a pair of A-molecules and the last term signifies the 
annihilation of a B-molecule. The function $\sigma(\vec R|\vec r,\vec r^\prime)$ or $\sigma^\prime(\vec r,\vec r^\prime|\vec R)$ 
denotes the reaction rate in each direction of (8.1). 
In physics, the Hamiltonian (8.2) describes a 
system of interacting bosons of two species. 
The equilibrium state of the fluorescence-diffusion-reaction process will be
analogous to the ground state which carries a Bose condensate. The proportion of each species in the condensate is 
determined by the
mass-action law and the fluctuations are calculatable with well
developed field theoretic method.  In terms of the generating 
functional (4.2) and the functional derivative (4.4), we 
are able to calculate various correlation functions of the photon 
counting histogram in the presence of chemical reactions using techniques 
developed in quantum field theory.  This approach is expected to be more 
systematic than the conventional reaction kinetics. 

\bigskip
\noindent{\bf{Acknowlegment}}\bigskip

The works of N.L. Goddard, 
G. Bonnet and A. Libchaber are supported in part by 
Mathers Foundation and the Burrough-Welcome Fund. The work of H. C. 
Ren is supported in part by US Department of Energy 
under the contract DE-FG02-91ER40651-TASK B.  We would also like to 
thank David Mauzerall for his discussions and suggestions.

\bigskip
\noindent{\bf{Appendix A}}
\bigskip

\noindent{\it{A.1 The derivation of the formulae for distribution 
function of the time delay histogram}}.

\bigskip
To derive the $\Delta t$ distribution (the distribution function of 
the time interval between two successive photon emissions), we start 
with the case with finite time bins ($\epsilon$ is sufficiently 
small that the probability of more than one photons within a bin can 
be ignored.) and look for the probability of one photon event in 
$n^{th}$-bin, one photon event in $(n+l)^{th}$ bin with $\Delta 
t=(l-1)\epsilon$ and no photon in the bins between them. Up to a 
normalization constant, the probability is 
$$\rho_{nl}={\rm{const.}}\sum_{m_1,...,m_{n-1},m_{n+l+1},...,m_{\cal 
N}}P_{m_1...m_{n-1}10...,01m_{n+l+1}...m_{\cal N}}$$
$$={\partial\over\partial z_n}{\partial\over\partial z_{n+l}}{\cal 
G}(1,...,1,z_n,0,...,0,z_{n+l},1,...,1).\eqno(A.1)$$ By summing 
over $n$ with a fixed $l$, we find the probability of successive 
$l-1$ empty bins $$\rho_l=\sum_n\rho_{nl}.\eqno(A.2)$$Taking the 
limit of infinitesimal bins, i.e., $\epsilon\to 0$ at fixed $T$ and 
$\Delta t$, we obtain the desired distribution function (4.7).

\bigskip\noindent{\it{A.2 Detector effect}}

\bigskip Consider the 
case of a single time bin, i.e.${\cal N}=1$, we have $${\cal 
G}(z)=\sum_{m=0}^\infty P_mz^m\eqno(A.3)$$ and $${\cal 
G}_{\rm{eff.}}(z)=\sum_{m=0}^\infty P_m^{\rm{eff.}}z^m.\eqno(A.4)$$
Under the assumption that each photon counting by the detector is 
statistically independent of others, the probability of detecting $m$ 
photons out of $n$ incident photons is $${n!\over 
m!(n-m)!}\eta^m(1-\eta)^{n-m}\eqno(A.5)$$ with $\eta$ the detector 
efficiency. Therefore$$P_m^{\rm{eff.}}=\sum_{n=m}^\infty P_n{n!\over 
m!(n-m)!}\eta^m(1-\eta)^{n-m}.\eqno(A.6)$$ Substituting 
(A.6) into (A.4), we obtain that $${\cal 
G}_{\rm{eff.}}(z)=\sum_{m=0}^\infty\sum_{n=m}^\infty P_n {n!\over 
m!(n-m)!}\eta^m(1-\eta)^{n-m}z^m = \sum_{n=0}^\infty 
P_n\sum_{m=0}^n{n!\over m!(n-m)!}(\eta z)^m(1-\eta)^{n-m}$$ $$=\sum_{n=0}^\infty 
P_n(1-\eta+\eta z)^n ={\cal G}(1-\eta+\eta 
z).\eqno(A.7)$$ Following the same steps for each variable in the 
case with ${\cal N}>1$, we end up with (4.10) and (4.11). 

\bigskip\noindent{\bf{Appendix B}}\bigskip

The probability generating functional of the photon emission histogram from $N$ identical 
fluorescence molecules is modeled according to the following two 
principles:

1). The generating functional with $N$ nonreacting 
molecules $${\cal G}[z(t)]=g^N[z(t)]\eqno(B.1)$$ with $g[z(t)]$ the 
generating functional of one molecule. 

2). The generating functional of one molecule,$$g[z(t)]=\sum_{\cal 
C}P_{\cal C}g_{\cal C}[z(t)],\eqno(B.2)$$ where 
$g_{\cal C}[z(t)]$ denote the generating functional along a 
particular diffusion path, ${\cal C}$, and $P_{\cal C}$ the 
probability of the path.

Dividing the integrating time $T$ into ${\cal N}$ time bins with $t_0=0,...,t_n=n\epsilon,...,t_{\cal 
N}={\cal N}\epsilon=T$ and specifying a diffusion path ${\cal C}$ by 
the location of the molecule at each instant, $(t_0,...,t_n,...,T)$, 
i.e. $${\cal C}=\{\vec r_0,...,\vec r_n,...,\vec r_{\cal 
N}\}.\eqno(B.3)$$ For sufficiently small $\epsilon$, the probability 
of more than one photon in each bin may be ignored and we have
$$g_{\cal C}[z(t)]=\lim_{\epsilon\to 0}\prod_{n=0}^{{\cal N}-1}[1-\epsilon\lambda u(\vec r_n)+\epsilon\lambda z_nu(\vec r_n)]$$
$$=\lim_{\epsilon\to 0}e^{\lambda\epsilon\sum_{n=0}^{{\cal 
N}-1}(z_n-1)u(\vec r_n)}.\eqno(B.4)$$
The probability of the path ${\cal C}$ is entirely determined by diffusion. Since the probability for a 
molecule to diffuse from $\vec r_n$ at $t_n$ to $\vec r_{n+1}$ at 
$t_{n+1}$ is $$\Big({1\over 4\pi D\epsilon}\Big)^{3\over 2}d^3\vec 
r_{n+1}e^{-{(\vec r_{n+1}-\vec r_n)^2\over 4D\epsilon}},\eqno(B.5)$$
$$P_{\cal C}=\prod_{n=0}^{{\cal N}-1}\Big({1\over 4\pi D\epsilon}\Big)^{3\over 2}d^3\vec r_{n+1}e^{-{(\vec r_{n+1}-\vec r_n)^2\over 4D\epsilon}}.\eqno(B.6)$$
It follows from (B.2), (B.4) and (B.6) 
that $$g[z(t)]={1\over\Omega}\lim_{{\cal N}\to\infty}\Big({1\over 4D\pi\epsilon}\Big)^{3{\cal N}\over 2}\int\prod_{n=0}^{\cal N} d^3\vec r_ne^{-\epsilon\sum_nL_n},\eqno(B.7)$$
with $$L_n={1\over 4D}\Big({\vec r_{n+1}-{\vec r_n}\over\epsilon}\Big)^2+(1-z_n)\lambda u(\vec r_n),\eqno(B.8)$$ where we have taken the average of the 
intial location of the molecule, $\vec r_0$ over the volume of the 
solution. Mathematically, eq.(B.7) and (B.8) present a path 
integral of a quantum mechanical particle moving in an external 
potential $[1-z(t)]\lambda u(\vec r)$ in an imaginary time. A similar 
path integral has been used to describe fluorescence 
correlations.\cite{Enderlein96}
Following the standard procedure \cite{Feynman65}, we may cast (B.7) 
into the canonical form $$g[z(t)]={1\over\Omega}<|{\cal 
T}e^{-\int_0^Tdth(t)}|>\eqno(B.9)$$ where ${\cal T}$ is the time 
ordering operator,  $$h(t)=-D\vec\nabla^2+[1-z(t)]\lambda u(\vec 
r)\eqno(B.10)$$ is the analog of the quantum mechanical Hamiltonian 
operator and $|0)$ is the analog of a quantum mechanical state, 
whose wave function is $<\vec r|>=1$.

Finally, we would like to explain the operator 
${\cal T}$ in more detail, when acting on a product of operator 
functions of time, it arranges the order of these operators 
according to the descending order of their time arguments, i.e.
$${\cal 
T}O(t_1)O(t_2)...O(t_n)=O(t_{P_1})O(t_{P_2})...O(t_{P_n})\eqno(B.11)$$
with $P_1,P_2,...,P_n$ a permutation of $1,2,...,n$, such that 
$t_{P_1}\geq t_{P_2}\geq...\geq t_{P_n}$. This property, when applied 
to the Taylor exponential operator in (B.9), yields: $${\cal 
T}e^{-\int_0^Tdth(t)}=1-\int_0^Tdth(t)+\int_0^Tdt_2\int_0^{t_2}dt_1h(t_2)h(t_1)+...$$ 
$$+(-)^n\int_0^Tdt_n\int_0^{t_n}dt_{n-1}...\int_0^{t_2}dt_1h(t_n)h(t_{n-1})...h(t_1)+....\eqno(B.12)$$

\noindent{\bf{Appendix C}}\bigskip
\bigskip

For a time independent operator $A$ and a time dependent operator $B(t)$,
the following identity holds with the time ordering product:
$${\cal T}e^{-\int_0^td\tau[A+B(\tau)]}=e^{-tA}\Big[{\cal T}e^{-\int_0^Td\tau{\cal B}(\tau)}\Big]\eqno(C.1)$$ 
with ${\cal B}(\tau)=e^{\tau A}B(\tau)e^{-\tau A}$.  

The expansion of the functional (4.21) that generates all correlation coefficients 
for a non-reacting system follows from the identity (C.1) with 
$A=-D\nabla^2$ and $B=[1-z(t)]\lambda u(\vec r)$. We find
$${\cal F}[z(t)]=c\Big[\int_0^Tdt(1-z(t))<|u|>-\int_0^Tdt_2(1-z(t_2))\int_0^t2dt_1
(1-z(t_1))<|ue^{(t_2-t_1)D\nabla^2}u|>+...\Big].\eqno(C.2)$$ 
$$I=C_1(t)=c\lambda<|u|>,\eqno(C.3)$$ and
$$C_2(t,t^\prime)=c\lambda^2<|ue^{|t-t^\prime|D\nabla^2}u|>.\eqno(C.4)$$
This expansion is parallel to the perturbative expansion of the quantum mechanical 
analog.

The expansion in the diffusion constant
for a non-reacting system is obtained by applying the 
identity (C.1) with $A=-\lambda u$, $B=-D\nabla^2$ and $t=\Delta t$, 
and making a Taylor expansion of the second factor on the right hand 
side of (C.1).
$$nF(\tau|\zeta)=c\Big[<|1-e^{-\zeta\lambda\tau u}|>-\int_0^\tau ds<|e^{-\zeta\lambda(\tau-s)u}
D\nabla^2e^{-s\zeta\lambda\tau u}|>+...\Big].\eqno(C.5)$$
This expansion corresponds to the strong coupling expansion of the quantum mechanical 
analog. 

The calculation of the expectation value $<|...|>$ in (C.2) and (C.5) facilitated by switching 
between the coordinate and momentum representations of the quantum mechanical analg. The state 
$|>$ is the state of zero momentum and is normalized the the volume of the system.

\pagebreak

\section{Figures \& Captions}

\noindent 1. Schematic of a photon counting trace.  Traditional photon 
counting divides the total integration time into bins, counting the 
number that fall into each.  We propose a new type of counting based 
on the time delay between two successive photon counts.\\

\noindent 2. Schematic of the photon counting setup. OBJ=objective, 
DM=dichroic mirror, NF=Notch Filter, PH=Pinhole, APD=Avalanche 
Photodiode, 
CB=Counting Board, CO=Correlator Board\\

\noindent 3. The time-delay histograms for single (black) and double (grey) labeled 
ssDNA.  The resulting distributions from a single Poisson processes are 
shown by the straight lines for comparison.\\

\noindent 4. (A)large $\Delta t$ single fluor, (B)small $\Delta 
t$ single fluor, 
(C)large $\Delta t$ double fluor, (D)small $\Delta t$ double 
fluor.  
(Solid) Poisson, (short dash) linear, and (long dash) quadratic fits are shown for comparison in the 
short time limits.\\

\noindent 5. Q Parameters of ($\bullet$) single dye and ($\triangle$) double dyed 
aggregates.  A cartoon of each system is displayed next to the 
corresponding curve.
\\
\pagebreak

\pagebreak

\pagebreak

\pagebreak

\pagebreak

\end{document}